\documentclass[preprint,showpacs,showkeys,preprintnumbers,amsmath,amssymb]{revtex4}
\usepackage{graphicx,epsfig}
\usepackage{dcolumn}
\usepackage{bm}

\def\ut#1{\mathop{\vtop{\ialign{##\crcr
     $\hfil\displaystyle{#1}\hfil$\crcr\noalign
     {\kern1pt\nointerlineskip}\hbox{$\hfil\sim\hfil$}\crcr
     \noalign{\kern1pt}}}}}

\begin{document}

\preprint{}

\title{

Are signatures of anti-de-Sitter black hole at the Galactic Center?
}

\author{Alexander F. Zakharov$^{1,2,3,4,5}$}
\email{zakharov@itep.ru}
 \affiliation{$^{1}$North Carolina Central University, Durham, NC 27707,
 USA\\
  $^{2}$Institute of
Theoretical and Experimental Physics, Moscow, 117218, Russia\\
   $^{3}$Joint Institute for Nuclear Research, Dubna, 141980,  Russia\\
   $^{4}$Institute for Computer Aided Design of RAS,     123056, Moscow, Russia\\
   $^{5}$National Research Nuclear University (NRNU MEPHI), 115409,
   Moscow, Russia
}

\date{\today}

\begin{abstract}

Using Schwarzschild -- de-Sitter  (Kottler) metric we derive a
simple analytical relation between a shadow size and $\Lambda$-term.
Current observations of the smallest spot to evaluate shadow size at
the Galactic Center do not reach an accuracy comparable with
cosmological $\Lambda$-term $\sim 10^{-52}{\rm m}^{-2}$, however, if
in reality we have dark energy instead of a constant $\Lambda$-term
then dark energy may be a function depending on time and space and
it could be approximated with a local constant $\Lambda$-term near
the Galactic Center and it is important to introduce a procedure to
evaluate  the $\Lambda$-term. We suggest such a procedure based on a
black hole shadow evaluation. Surprisingly, current observational
estimates of shadows are in agreement
 with
anti-de-Sitter spacetimes corresponding to a negative $\Lambda$-term
which is about $-0.4\times 10^{-20}{\rm m}^{-2}$. A negative
$\Lambda$-term has been predicted in the framework of a some class
of multidimensional string models.

\end{abstract}


\maketitle

\section{Introduction}

First solutions corresponding to spherical symmetric black holes
were found \cite{Schwarzschild_16,Kottler_18} soon after discovery
of general relativity. Now the Kottler metric is called usually as
the Schwarzschild -- de-Sitter one because it consists of the
Schwarzschild  and the de-Sitter solutions for a suitable choice of
parameters  and one has the black hole and cosmological (de-Sitter)
metrics as corresponding limiting cases. We will not discuss the
topological structure of the solution was analyzed in details
\cite{Gibbons_77}.

Generally speaking, an appearance of the $\Lambda$-term (introduced
by Einstein \cite{Einstein_17}) in the geometrical part of the
Einstein equations are rather natural and one can find it in old
textbook and monographs published even before 1998 (see, for
instance, \cite{Tolman_34}) when it was established the
$\Lambda$-term presence in the conventional cosmological models
using SNIa data \cite{Riess_98} (the discovery was confirmed with
alternative techniques, see \cite{Cosmology_2014} and references
therein), however, perhaps the Gamow's story (or the fairy tale) on
the Einstein's statement about "the biggest blunder" \cite{Gamow_70}
concerning $\Lambda$-term in the gravitational field equations to
create a static cosmological model (see, for a discussion
\cite{Livio_13} and \cite{Weinstein_13}) significantly broke
intensive studies of all aspects of the $\Lambda$-term presence in
the Einstein's gravitational field equations.

If the $\Lambda$-term is putting in the stress-energy part of
Einstein's field equations it can be treated as an additional
component of energy (so-called dark energy introduced in
\cite{Turner_99}) and in principle the component can be slightly
different from constant $\Lambda$ and may be function of spacetime
coordinates. There are many theoretical and observational proposals
to investigate dark energy, including investigations of dark energy
accretion onto black holes \cite{Babichev_13}.

In spite of the fact that the Schwarzschild -- de-Sitter solutions
are known for many years, there is still a discussion about
observable signatures of the $\Lambda$-term in gravitational lensing
starting probably since the Islam's claim that a $\Lambda$-term
presence is observable \cite{Islam_83}, while Rindler \& Ishak
criticized this statement \cite{Rindler_07} (see, papers reflecting
different opinions concerning the issue \cite{GL_lambda} and
references therein). The most of these papers have no mistakes but
opposite sides use different approaches and there is no agreement in
definition of observable quantities and usually it is very hard to
evaluate accuracies of suggested approaches and contributions of
systematic errors.

In spite of the fact that black hole solutions of Einstein equations
are known for almost century there are not too many astrophysical
examples where one really need a strong gravitational field
approximation but not small relativistic corrections to a Newtonian
gravitational field. One of the most important option to test a
gravity in the strong field approximation is analysis of
relativistic line shape as it was shown in \cite{Fabian_89}. Later
on, such signatures of the Fe $K\alpha$-line have been found in the
active galaxy MCG-6-30-15 \cite{Tanaka_95}. Analyzing the spectral
line shape the authors concluded the emission region is so close to
the black hole horizon that one has to use Kerr metric approximation
to fit observational data \cite{Tanaka_95}. Results of our
simulations of iron $K_\alpha$ line formation are given in
\cite{Zakharov_99} (where we used our approach \cite{Zakharov_94}),
see also \cite{Fabian_10} for  more recent reviews of the subject.

There are two basic observational techniques to investigate a
gravitational potential at the Galactic Center, namely, a)
monitoring the orbits of bright stars near the Galactic Center to
reconstruct a gravitational potential \cite{Ghez_00} (see also a
discussion about an opportunity to evaluate black hole dark matter
parameters in \cite{Zakharov_07} and an opportunity to constrain
some class of an alternative theory of gravity \cite{Dusko_PRD_12});
b) in mm-band with VLBI-technique measuring a size and a shape of
shadows around black hole giving an alternative possibility to
evaluate black hole parameters. Doeleman et al. evaluated a size of
the smallest spot for the black hole at the Galactic Center
 with VLBI technique in mm-band \cite{Doeleman_08} (see also constraints done from previous
observations \cite{Shen_05}). Theoretical
 studies showed that the size of the smallest spot near a black hole
 practically coincides  with shadow size because the spot is
 the envelope of the shadow \cite{Falcke00,ZNDI05,ZNDI05b}.
As it was shown \cite{ZNDI05,ZNDI05b}, measurements of the shadow
size around the black hole may help to evaluate parameters of black
hole metric \footnote{One of the first calculations of shapes of
orbits visible by  a distant observer have been done in
\cite{Cunningham_73}. An apparent shape of Kerr black hole was
discussed in \cite{Bardeen_73} (see also a very similar picture in
monograph \cite{chandra}). Later on, the apparent shapes of black
holes are called shadows \cite{Falcke00}.}. Sizes and shapes of
shadows are calculated for different types of black holes and
gravitational lensing in strong gravitational field has been
analyzed in a number of papers \cite{Shadows}.

 We derive an analytic expression for the
black hole shadow size as a function of the $\Lambda$-term. The
result shows that there is an $\Lambda$-term impact on observational
quantities for  gravitational lensing because shadow formation is a
clear illustration of gravitational focusing.

 We conclude that
observational data concerning shadow size evaluation for the
Galactic Center are consistent with significant negative
$\Lambda$-terms which play important role in string theory
\cite{Horowitz_98} (see also, \cite{Susskind_05}).

\section{Basic Equations}

There is the Kerr -- Newman -- de-Sitter solution \cite{Carter_73}
which a generalization of the Kerr -- Newman solution with the
$\Lambda$-term. Carter separated variables and found a complete set
of first integrals for this case \cite{Carter_73} similarly to the
Kerr -- Newman case \cite{Carter_68}.

 The expression for the Kottler (Schwarzschild --
de-Sitter) metric in natural units ($G=c=1$) has the form
\cite{Kottler_18,Carter_73,Stuchlik_83}, we
\begin{equation}
  ds^{2}=-\left(1-\frac{2M}{r}-\frac{1}{3}\Lambda r^2\right)dt^{2}+\left(1-\frac{2M}{r}-\frac{1}{3}\Lambda r^2\right)^{-1}dr^{2}+
r^{2}(d{\theta}^{2}+{\sin}^{2}\theta d{\phi}^{2}).
\label{Lambda_0}
\end{equation}
where $M$ is the mass of the black hole and we use the conventional
nomination for the $\Lambda$-term.

We will remind the horizon structure
 for $\Lambda < 0$. Really, we have  for the metric (\ref{Lambda_0})
 (see, for instance \cite{Stuchlik_83})
\begin{equation}
r_h=-\left[-\frac{3M}{\Lambda}+\left(\frac{9M^2}{\Lambda^2}-\frac{1}{\Lambda^3}\right)^{1/2}\right]+
\left[-\frac{3M}{\Lambda}-\left(\frac{9M^2}{\Lambda^2}-\frac{1}{\Lambda^3}\right)^{1/2}\right].
\label{Lambda_0a}
\end{equation}
while for $\Lambda > 0$ we have three apparent singularities
corresponding to vanishing $g_{00}$ component of the metric tensor
\begin{equation}
r_u=-2\Lambda^{-1/2}\cos\left(\frac{1}{3} \psi\right)
\label{Lambda_0b}
\end{equation}
\begin{equation}
r_h=2\Lambda^{-1/2}\cos\left(\frac{1}{3}(\pi+\psi)\right),
\label{Lambda_0c}
\end{equation}
\begin{equation}
r_c=-2\Lambda^{-1/2}\cos\left(\frac{1}{3}(\pi-\psi)\right),
\label{Lambda_0d}
\end{equation}
where
\begin{equation}
\psi=\arccos(3M \Lambda^{1/2}),
\label{Lambda_0e}
\end{equation}
$r_h, r_u, r_c$ are measuring in $M$ units.

Applying the Hamilton--Jacobi method to the problem of 
geodesics in the Kottler metric, 
the motion of a test particle in the $r$-coordinate can be described
by following equation (see, for example,
\cite{Carter_73,Stuchlik_83})
\begin{eqnarray}
    r^{4}(dr/d\lambda)^{2}=R(r),\label{Lambda 1}
\end{eqnarray}
where $\lambda$ is the affine parameter \cite{MTW,Wald_84} and for
the polynomial $R(r)$ we have \cite{Carter_73,Stuchlik_83}
\begin{eqnarray}
 R(r)=
 r\left[
 \frac{1}{3} \Lambda m^2 r^5-(E^2-m^2+\frac{1}{3} \Lambda L^2) r^3+2m^2 M r^2-L^2 r+2ML^2
 \right],    \label{Lambda_2}
\end{eqnarray}
Here, the constants $m, E$ and $L$  are associated with the
particle, i.e. $m$ is its mass, $E$ is energy at infinity and $L$ is
its angular momentum at infinity.

    We will consider the motion of photons $(m=0)$.
In this case, the expression for the polynomial $R(r)$ takes the
form
\begin{eqnarray}
R(r)=r\left[(E^2+\frac{1}{3}\Lambda L^2)r^3 - L^2 r +2 L^2\right].
\label{Lambda_3}
\end{eqnarray}

     Depending on the multiplicities of the roots of the
polynomial $R(r)$,  we can have three types of photon motion in the
$r$ - coordinate \citep{Zakharov_86}. If an observer is located at
infinity for $\Lambda < 0$ or at $r_o$ ($r_h < r_o < r_c$), we have:

 (1) if the polynomial $R(r)$
has no roots  for $r\geq r_{h}$,  a test particle is captured by the
black hole;

(2) if $R(r)$ has roots  and $(\partial{R}/\partial{r})(r_{max})\neq
0$  with $r_{max} > r_h$ ($r_{max}$ is the maximal root), a particle
is scattered after approaching the black hole;

(3) if  $R(r)$ has a root  and $R(r_{max}) = (\partial{R}/
\partial{r})(r_{max})=0$,
the particle now takes an infinite proper time to approach the
surface $r = const$.

    If we are considering a photon ($\mu = 0$), its motion in the $r$-coordinate depends
on the root multiplicity of the polynomial $\hat{R}(\hat{r})$
\begin{eqnarray}
\hat{R}(\hat{r})={R(r)}/({M^{4}E^{2}})=\hat{r}\left[(1+\frac{1}{3}\hat{\Lambda}\xi^2)
-\xi^{2}{\hat{r}}^{2}+2{\xi}^{2}\right],\label{Lambda_4}
\end{eqnarray}
where $\hat{r}=r/M, \xi=L/(ME)$ and $\hat{\Lambda}=\Lambda M^2.$
Below we do not write $\wedge$ symbol for these quantities.

\section{Derivation of shadow size as a cosmological constant}

    Let us consider the problem of the capture cross section of
 a photon by a black hole. It is clear that the critical value of the
 impact parameter for a photon to be captured by a  black hole
depends on the root multiplicity of the polynomial $R(r)$. This
requirement is equivalent to the  vanishing discriminant condition
\cite{Kostrikin_82}. Earlier, to find the critical value of impact
parameter for Schwarzschild and Reisner -- Nordstr\"om metrics the
condition has been used for corresponding cubic and quartic
equations \cite{Zakharov_88,Zakharov_91,Zakharov_94b}. In
particular, it was shown that for these cases the vanishing
discriminant condition approach is more powerful in comparison with
the procedure excluding $r_{max}$ from the following system
\begin{eqnarray}
R(r_{max}) = 0, \quad  \\
\label{Lambda_4a}
 \dfrac{\partial{R}}{\partial{r}}(r_{max})=0,
\label{Lambda_4b}
\end{eqnarray}
as it was done, for example,  by Chandrasekhar \cite{chandra} (and
earlier by Darwin \cite{Darwin_58}) to solve similar problems,
because $r_{max}$ is automatically excluded in the condition for
vanishing discriminant.

Introducing the notation $\alpha=\dfrac{\xi^2}{1+\frac{1}{3}\Lambda
\xi^2}$, we obtain
\begin{eqnarray}
    f(r)=R(r)/(1+\frac{1}{3}\Lambda \xi^2)=r^{3}-\alpha r^{2}+2\alpha.\label{Lambda_5}
\end{eqnarray}

We remind the discriminant definition for cubic equation. If we
consider an reduced polynomial $g(x)$ with the third degree
\begin{eqnarray}
g(x) =  x^3+ax+b, \label{Lambda_6}
\end{eqnarray}
then the discriminant is \cite{Kostrikin_82}
\begin{eqnarray}
D =  -4a^3-27b^2, \label{Lambda_7}
\end{eqnarray}
Therefore, for the polynomial (\ref{Lambda_5}) we have the following
condition for the multiple roots
\begin{eqnarray}
D =  4\alpha^3-108 \alpha^2, \label{Lambda_7a}
\end{eqnarray}
or $\alpha=27$, so
\begin{eqnarray}
\xi_{\rm cr}^2=\frac{27}{1-9\Lambda}. \label{Lambda_8}
\end{eqnarray}
To find a radius of photon unstable orbit we will find the multiple
root $r_{\rm cr}$ of the polynomial (\ref{Lambda_5}) substituting
$\xi_{\rm cr}$ in the relation. It is easy to see that $r_{\rm
cr}=3$.  Similar results can be obtained with analysis of
corresponding potentials \cite{Stuchlik_83} as it was described
earlier. For    $\Lambda=0$ one obtains the well-known result for
Schwarzschild metric \cite{MTW,Wald_84,Lightman_75}. However, the
vanishing discriminant condition allows immediately eliminate the
radial coordinate in equations for critical values of the
parameters.

As it was explained in \cite{ZNDI05b} this leads to the formation of
shadows  described by the critical value of $\xi_{cr}$ or, in other
words, in the spherically symmetric case, shadows are circles with
radii $\xi_{\rm cr}$. Therefore, by measuring the shadow size, one
could evaluate the cosmological constant in  black hole in units
$M^{-2}$.

\section{Consequences}

\subsection{Observational constraints on the $\Lambda$-term
at the Galactic Center}

\begin{figure}[th!]
\begin{center}
\includegraphics[width=10.5cm]{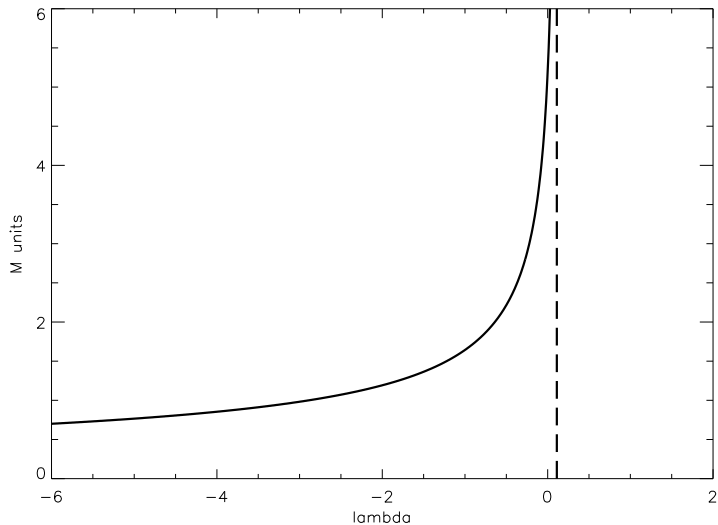}
\end{center}
\caption{Shadow (mirage) radius (solid line) in $M$ units as a
function of dimensionless
 $\Lambda$. The critical value $\Lambda=1/9$ is shown with the dashed vertical line.}
 \label{Fig1}
\end{figure}

As one can see from Eq.  (\ref{Lambda_8}) shadows disappear for
$\Lambda > 1/9$, and there exist for $\Lambda < 1/9$ and for
positive $\Lambda$ its presence decrease shadow dimension while for
negative $\Lambda$ we have an opposite tendency (see, Fig.
\ref{Fig1}).

If we adopt the distance toward the Galactic Center $d_*=8.3(\pm
0.4)$~kpc (or $d_*=8.35(\pm 0.15)$~kpc \cite{Reid_14}) and mass of
the black hole $M_{BH}=4.3(\pm 0.4) \times 10^6 M_\odot$
\cite{Gillessen_09,Falcke_13} (a significant part of black hole mass
uncertainty is connected with a distance determination uncertainty
\cite{Falcke_13}), then we have the angle 10.45~$\mu as$ for the
corresponding Schwarzschild radius $R_g=2.95*
\dfrac{M_{BH}}{M_\odot}*10^5$~cm roughly with 10\% uncertainty of
black hole mass and distance estimations, so a shadow size for the
Schwarzschild black hole is around 53~$\mu as$.


A couple of years ago Doeleman et al. \cite{Doeleman_08} claimed
that intrinsic diameter of Sgr $A^*$ is $37^{+16}_{-10}~\mu as$ at
the $3~ \sigma$ confidence level. If we believe in GR and the
central object is a black hole, then we have to conclude that a
shadow is practically coincides with the intrinsic diameter, so in
spite of the fact that a Schwarzschild black hole is marginally
consistent with observations, a Schwarzschild -- anti-de-Sitter
black hole provides much better fit of a shadow size. Later on, an
accuracy of intrinsic size measurements was significantly improved,
so  Fish et al. \cite{Fish_11} gave $41.3^{+5.4}_{-4.3}~\mu as$ (at
3~$\sigma$ level) on day 95, $44.4^{+3.0}_{-3.0}~\mu as$ on day 96
and $42.6^{+3.1}_{-2.9}~\mu as$ on day 97. If we adopt a shadow size
as $43~\mu as$, we can conclude that the $\Lambda$-term is about
$-0.4\times 10^{-20}m^{-2}$, so we have signatures of Schwarzschild
-- anti-de-Sitter black hole.

The absolute value the $\Lambda$-term is many orders of magnitude
higher than the mean value expected from cosmological data, however,
an enormous progress of observational facilities gives a hope that a
precision of the estimates of dark energy component from this kind
of data will be significantly approved very soon, moreover, for
black holes with higher masses the $\Lambda$ estimate can be more
accurate. So, since $\hat{\Lambda}$ is evaluating in $M^{-2}$ units
one can significantly improve an accuracy of $\Lambda$ estimates
increasing a black hole mass. For instance, $M_{M87}$ is higher
$M_{Sgr A*}$ more than in $10^3$ therefore $\Lambda$ estimates may
be at least in $10^6$ times better with the same observational
facilities.

The black hole in the elliptical galaxy M87 looks also perspective
to evaluate shadow size \cite{Doeleman_12} (probably  even its shape
in the future to estimate a black hole spin) because the distance
toward the galaxy is $16 \pm 0.6 $~Mpc \cite{Blakeslee_09}, black
hole mass is $M_{M87}=(6.2\pm 0.4) \times 10^9 M_\odot$
\cite{Gerbhardt_11}, so that an angle $(7.3 \pm 0.5) \mu as$
corresponds to the Schwarzschild radius \cite{Doeleman_12}, so the
angle is comparable with the corresponding value considered earlier
for our Galactic Center case. Recently, it was reported that
smallest shadow size is $5.5 \pm 0.4  R_{SCH}$ with $1~\sigma$
errors  (where $R_{SCH}=2 G M_{M87}/c^2$) \cite{Doeleman_12}, so
that at the moment the shadow size is consistent with the
Schwarzschild metric for the object.

An existence of cosmological black holes with masses around
$10^{13}-10^{17}M_\odot$ has been discussed \cite{Stornaiolo_02}. As
it was explained earlier in this case an analysis of shadows could
give a much better estimates of $\Lambda$-term around such objects
with the same angular accuracy of the shadow evaluation.

\section{Conclusions}

Solar system effects caused by $\Lambda$-term analyzed in details
\cite{Kagramanova_06} and the most sensitive effect is perihelion
shift found in \cite{Kerr_03} and the Pioneer anomaly if exists
\cite{Turyshev_09}. However, as the Pioneer anomaly example shows it
is not a simple problem to evaluate systematic errors and
non-gravity effects in such studies of gravitational field in Solar
system.

Based on observations \cite{Doeleman_08,Fish_11} one can say that
for the Schwarzschild black hole model we have tensions between
evaluations of black hole mass done with observations of bright star
orbits near the Galactic Center and the evaluated shadow size. To
reduce tensions between estimates of the black hole mass and the
intrinsic size measurements, one can use the Schwarzschild --
anti-de-Sitter metric with a significant negative $\Lambda$-term
around $-0.4\times 10^{-20}m^{-2}$. However, in spite of the
signatures of the negative $\Lambda$-term, one can not claim that
anti-de-Sitter black hole has been discovered in the Galactic Center
because the issue needs further observational studies and
confirmations.

Recent estimates of the smallest structure in the M87 published in
paper \cite{Doeleman_12} do not need an introduction of $\Lambda$
-term (positive or negative) to fit observational data because sizes
of the smallest spot near the black hole at the object are still
consistent with the shadow size evaluated for the Schwarzschild
metric.

There is a clear impact of $\Lambda$-term on gravitational lensing
because a shadow formation is gravitational lensing phenomenon in a
strong gravitational field and shadow dimensions are observable in
spite of the fact that at the moment an accuracy of such
measurements is not good enough to evaluate a $\Lambda$-term
comparable with the cosmological constant.

\begin{acknowledgments}

The author thanks D. Borka, V. Borka Jovanovi\'c, F. De Paolis, G.
Ingrosso, P. Jovanovi\'c, S. M. Kopeikin,  A. A. Nucita, B. Vlahovic
for useful discussions. The author acknowledges also A. Broderick
and C. L\"ammerzahl for conversations at GR-20/Amaldi-10 in Warsaw.

 The work was supported in part by the NSF
(HRD-0833184) and NASA (NNX09AV07A) at NASA CADRE and NSF CREST
(NCCU, Durham, NC, USA) and RFBR 14-02-00754a at ICAD of RAS
(Moscow).
\end{acknowledgments}


\begin{thebibliography}{}

\bibitem[\protect\citeauthoryear{Schwarzschild}{1916}]{Schwarzschild_16}
K. Schwarzschild, Sitzunsber. Preuss. Akad. Wiss. Berlin (Math.
Phys.) 189, (1916); see, English translation in
arxiv.org/physics/9905030v1.

\bibitem[\protect\citeauthoryear{Kottler}{1918}]{Kottler_18}
F. Kottler, Annalen der Physik, {\bf 361}, 401 (1918).

\bibitem[\protect\citeauthoryear{Gibbons \& Hawking}{1977}]{Gibbons_77}
G. W. Gibbons, S.W. Hawking,  Phys. Rev. D  {\bf 15}, 2738 (1977);

R. Bousso, arxive:hep-ph/0205177v2.

\bibitem[\protect\citeauthoryear{Einstein}{1917}]{Einstein_17}
A. Einstein, Sitzungsberichte der Kuniglich Preussischen Akademie
der Wissenschaften (Berlin),  142 (1917).

\bibitem[\protect\citeauthoryear{Gamow}{1970}]{Gamow_70}
 G. Gamow,  My World Line: An Informal Biography,  (New York, Viking Press, 1970).

\bibitem[\protect\citeauthoryear{Livio}{2013}]{Livio_13}
M. Livio, Brilliant Blunders: From Darwin to Einstein - Colossal
Mistakes by Great Scientists That Changed Our Understanding of Life
and the Universe (New York: Simon \& Schuster, 2013).

\bibitem[\protect\citeauthoryear{Weinstein}{2013}]{Weinstein_13}
G. Weinstein, arxiv.org/1310.1033v1[physics.hist-ph].

\bibitem[\protect\citeauthoryear{Tolman}{1934}]{Tolman_34}
R.C. Tolman, Relativity, Thermodynamics and Cosmology (Oxford,
Clarendon Press, 1934).


\bibitem[\protect\citeauthoryear{Riess et al.}{1998}]{Riess_98}
 A.G. Riess  et al., Astron. J. {\bf 116}, 1009
(1998);

B. P. Schmidt et al. Astrophys. J. {\bf 507}, 46 (1998);

S. Perlmutter et al., Astrophys. J. {\bf 517}, 565 (1999);
A.G.
Riess  et al., Astron. J. {\bf 560}, 49 (2001).

\bibitem[\protect\citeauthoryear{Cosmology}{2014}]{Cosmology_2014}


M. Tegmark et al., Phys. Rev. D {\bf 69}, 103501 (2004);

C. L. Bennett  et.al.,  Astrophys. J. Suppl. {\bf 208}, 20B (2013);

G.F. Hinshaw et.al.,  Astrophys. J. Suppl. {\bf 208}, 19H (2013);

O. Lahav, A. R. Liddle, arXiv:1401.1389 [astro-ph.CO];

Planck Collaboration: P. A. R. Ade et al.,
arxiv.org/1303.5076v3[astro-CO];

Planck Collaboration: P. A. R. Ade et al., Astron. \& Astrophys.,
{\bf 566},   A54 (2014);

A.G. Kim, et al. Astropart. Phys., http://dx.doi.org/10.1016/
j.astropartphys.2014.05.007 (2014).

\bibitem[Turner(1999)]{Turner_99}
M.S.  Turner, In {\em The Third Stromlo Symposium: The Galactic
Halo}, eds. B.K. Gibson, T.S. Axelrod \& M.E. Putman,  ASP
Conference Series {\bf 165},  p. 431 (1999).


\bibitem[\protect\citeauthoryear{Babichev et al.}{2013}]{Babichev_13}
E. O. Babichev, V. I. Dokuchaev, Yu. N. Eroshenko, Physics - Uspekhi
{\bf 56}, 1155 (2013).


\bibitem[\protect\citeauthoryear{Islam}{1983}]{Islam_83}
J. N. Islam,  Phys. Lett. A  {\bf 97}, 239 (1983).

\bibitem[\protect\citeauthoryear{Rindler \& Ishak}{2007}]{Rindler_07}
W. Rindler, M. Ishak, Phys. Rev. D {\bf 76}, 043006 (2007).

\bibitem[\protect\citeauthoryear{}{}]{GL_lambda}
E. Wright, arxiv:astro-ph/9805292v1;

M. Sereno, Phys. Rev. D {\bf 77}, 043004 (2008);


 A. Bhadra, S. Biswas and K. Sarkar, Phys. Rev. D {\bf 82}, 063003
(2010);

T. Schucker, Gen. Rel. Grav. {\bf 41}, 1595, (2009);

R. Kantowski, B. Chen and X. Dai, Astrophys. J. {\bf 718}, 913
(2010);

 M. Park, Phys. Rev. D {\bf 78}, 023014 (2008);

H. Arakida and M. Kasai, Phys. Rev. D {\bf 85}, 023006 (2012);

I. B. Khriplovich and A. A. Pomeransky, Int. J. of Mod. Phys. D {\bf
17}, 2255 (2008);

M. Ishak, W. Rindler, J. Dossett, J. Moldenhauer and C. Allison,
Mon. Not. R. Astro. Soc. {\bf 388}, 1279, (2008);

M. Ishak, Phys. Rev. D {\bf 78}, 103006 (2008);

M. Sereno, Phys. Rev. D {\bf 78}, 083003 (2008);

E. Hackmann and C. L\"ammerzahl, Phys. Rev. D {\bf 78} 024035
(2008);

E. Hackmann and C. L\"ammerzahl, Phys. Rev. Letters {\bf 100},
171101 (2008);


 M. Sereno, Phys. Rev. Letters {\bf
102}, 021301 (2009);


 F. Simpson, J. A. Peacock and A. F. Heavens, Mon. Not. R. Astro.
Soc. {\bf 402}, 2009, (2010);

M. Ishak, W. Rindler and J. Dossett, Mon. Not. R. Astro. Soc. {\bf
403}, 2152, (2010);

 M. Ishak and W. Rindler, Gen. Rel. Grav. {\bf 42},  2247,
(2010);

T. Biressa and J. A. de Freitas Pacheco, Gen. Rel. Grav. {\bf 43},
2649, (2011);

D. Lebedev, K. Lake, arxiv:1308.4931v1[gr-qc].

%
%
\bibitem{Fabian_89}
 A. C. Fabian, M.J. Rees, L. Stella, N. E. White, Month. Not. R. Astron. Soc., {\bf 238}, 729
 (1989);
L. Stella, Nature, {\bf 344}, 747 (1990);

{ A. Laor,  Astrophys. J.  {\bf 376}, 90 (1991)};

 G. Matt, G. C. Perola,
L. Stella,    Astron. \& Astrophys., {\bf 267}, 643 (1993).

\bibitem{Tanaka_95}
Y. Tanaka, K. Nandra, A. C. Fabian et al. Nature {\bf 375}, 659
(1995).
%
%
\bibitem[\protect\citeauthoryear{Zakharov \& Repin}{1999}]{Zakharov_99}
A.F.  Zakharov, S.V. Repin, Astron. Rep., {\bf 43}, 705 (1999);

 A.F.  Zakharov, S.V. Repin, Astron. Rep., {\bf 46}, 360 (2002);
%


  A.F.  Zakharov, S.V. Repin, Astron. \& Astrophys., {\bf 406}, 7 (2003);
%

%

  A.F.  Zakharov, S.V. Repin,  Astron. Rep., {\bf 47}, 733 (2003);

  A.F.  Zakharov, S.V. Repin,
   Nuovo Cimento, {\bf 118B}, 1193 (2003);
%
%
%
%
%

 A.F.  Zakharov, N.S.  Kardashev, V. N. Lukash, S.V. Repin,
   Month. Not. R. Astron. Soc., {\bf 342}, 1325 (2003);

\bibitem[\protect\citeauthoryear{Zakharov et al.}{2004}]{ZMB04}
A.F.  Zakharov, Z.  Ma, Y. Bao,
    New Astronomy, {\bf 9}, 663 (2004).



 A.F. Zakharov and S.V.  Repin,    Advances in Space Res.  {\bf 34},
2544 (2004);

A.F. Zakharov and S.V.  Repin, {Mem. S. A. It. della Supplementi,}
{\bf 7}, 60 (2005);

 A.F. Zakharov and S.V.  Repin,
{New Astron.}, {\bf 11}, 405 (2006);

A.F. Zakharov, {Phys. of Atom. Nucl.}, {\bf 70}, 159 (2007).

\bibitem[\protect\citeauthoryear{ Zakharov}{1994}]{Zakharov_94}
 A.F.  Zakharov, 1994, Month. Not. R. Astron. Soc.,  {\bf 269}, 283 (1994);

A.F.   Zakharov,  
in {\it Proceedings of 17th Texas Symposium on Relativistic
Astrophysics}, Ann. NY Academy of Sciences,  {\bf 759}, 550 (1995).
%
\bibitem{Fabian_10}
A.C. Fabian, R.R. Ross,  Space Sci. Rev. {\bf 157}, 167 (2010).

P. Jovanovi\'{c}, New Astron. Rev., {\bf 56}, 37 (2012).






\bibitem[\protect\citeauthoryear{Ghez et al.}{2000}]{Ghez_00}
A. M. Ghez, M. Morris, E.E. Becklin, A. Tanner, T. Kremenek, Nature
{\bf 407}, 349 (2000);

G. F. Rubilar and A. Eckart, Astron. Astrophys. {\bf 374}, 95
(2001);

 R. Sch\"odel, T.
Ott, R. Genzel, et al., Nature {\bf 419}, 694 (2002);

A.M. Ghez et al.,  2003, Astrophys. J. Lett. {\bf 586}, L127 (2003);

A.M. Ghez et al.,  Astrophys. J. Lett. {\bf 601}, L159 (2004);

A.M. Ghez et al.,  Astrophys. J. {\bf 620}, 744 (2005);

R. Genzel, R. Sch\"odel, T. Ott  et al.,   Nature {\bf 425}, 934
(2003);

 N. N. Weinberg, M.
Milosavljevi\'c, and A. M. Ghez, Astrophys. J. {\bf 622}, 878
(2005);

G. S. Adkins and J. McDonnell, Phys. Rev. D {\bf 75}, 082001 (2007);

S. Gillessen, F. Eisenhauer, S. Trippe, T. Alexander, R. Genzel, F.
Martins, and T. Ott, Astrophys. J. {\bf 692}, 1075 (2009);

S. Gillessen, R. Genzel, T. K. Fritz et al., Nature {\bf 481}, 51
(2012);

 L. Meyer, A. M.
Ghez, R. Sch\"odel, et al., Science {\bf 338}, 84 (2012);

{ M. R. Morris, L. Meyer,  A. M. Ghez, Research  Astron. Astrophys.
{\bf 12}, 995 (2012). }




\bibitem{Zakharov_07}
A. F. Zakharov, A. A. Nucita, F. De Paolis, and G. Ingrosso, Phys.
Rev. D {\bf 76}, 062001 (2007);

 A.F. Zakharov, S. Capozziello, F. De
Paolis, G. Ingrosso, A.A. Nucita, 
{Space Sci. Rev.} {\bf 48}, 301 (2009).

\bibitem{Dusko_PRD_12}
D. Borka,  P. Jovanovic,  V. Borka Jovanovic  and A. F. Zakharov,
{Phys. Rev. D}  {\bf 85}, 124004 (2012);

{ D. Borka, P. Jovanovi\'c, V. Borka Jovanovi\'c and A.F.
Zakharov, 
J.  Cosm. and Astropart. Phys.  {\bf 11}, 050 (2013).






%


\bibitem[\protect\citeauthoryear{Doeleman et al.}{2008}]{Doeleman_08}
 S.S. Doeleman  et al.,  {Nature}, {\bf  455}, 78 (2008).

 \bibitem[\protect\citeauthoryear{Shen et al.}{2005}]{Shen_05}
 Z. Q. Shen  et al.,  {Nature}, {\bf  438}, 62 (2005).

\bibitem[\protect\citeauthoryear{Falcke, Melia \& Agol}{2000}]{Falcke00}
H. Falcke, F. Melia, E. Agol,  Astrophys. J.  {\bf 528}, L13 (2000);

F. Melia, \& H. Falcke,     Annual Rev. Astron. \& Astrophys., {\bf
39}, 309 (2001).



\bibitem[\protect\citeauthoryear{Zakharov et al.}{2005}]{ZNDI05}
A.F. Zakharov, A.A. Nucita, F. De Paolis, G. Ingrosso,   New
Astronomy, {\bf 10}, 479 (2005);

 A.F. Zakharov, A.A. Nucita, F. De
Paolis, G. Ingrosso,  In: {\it Dark matter in astro- and particle
physics. Proc. the International Conference DARK 2004}, ed. H.-V.
Klapdor-Kleingrothaus, R. Arnowitt,  p. 77 (Berlin: Springer, 2005);

K. Hioki,  K.I. Maeda, Phys. Rev. D {\bf 80}, 024042 (2009).

\bibitem[\protect\citeauthoryear{Zakharov et al.}{2005}]{ZNDI05b}
A.F. Zakharov, A.A. Nucita, F. De Paolis, G. Ingrosso, Astron. \&
Astrophys. {\bf 10}, 795 (2005);

A.F. Zakharov, A.A. Nucita, F. De Paolis, G. Ingrosso, New Astron.
Rev., {\bf 56}, 64 (2012).

\bibitem[\protect\citeauthoryear{Cunningham \& Bardeen}{1973}]{Cunningham_73}
C.T. Cunningham,  J. M. Bardeen,
 Astrophys. J.  {\bf 183}, 237 (1973).


\bibitem[\protect\citeauthoryear{Luminet}{1979}]{Luminet_79}
J-P. Luminet, Astron. \&
 Astrophys. J.  {\bf 75}, 228 (1979);

J-P. Luminet, {\it Black Holes},  (Cambridge University Press,
1992).

\bibitem[\protect\citeauthoryear{Bardeen}{1973}]{Bardeen_73}
 J. M. Bardeen, in {\it Black Holes, Les Astres Occlus}, ed. by C.
de Witt, B.S. de Witt,
 p.~216 (Gordon and Breach Science Publishers, 1973).}

\bibitem[\protect\citeauthoryear{Chandrasekhar}{1983}]{chandra}
  S. Chandrasekhar, {\it Mathematical Theory of Black Holes},
   (Cla\-ren\-don Press, Oxford, 1983).

\bibitem[\protect\citeauthoryear{Shadows}{2014}]{Shadows}
{

 K. S. Virbhadra,  G. F. R. Ellis, Phys. Rev. D {\bf 65}, 103004 (2002);

E. F. Eiroa, Phys. Rev. D {\bf 73}, 043002 (2006);


A. Lobanov and J. A. Zensus, in {\it Exploring the Cosmic Frontier:
Astrophysical Instruments for the 21st Century}, ESO Astrophysics
Symposia, European Southern Observatory series, ed. A. P. Lobanov,
J. A. Zensus, C. Cesarsky and Ph. J. Diamond, p. 147,
(Springer-Verlag, Berlin and Heidelberg, 2007):
arxiv.org/astro-ph/0606143;


S.-M. Wu and T.-G. Wang, Month. Not. R. Astron. Soc., {\bf 378}, 841
(2007);


R. Takahashi, Month. Not. R. Astron. Soc., {\bf 382}, 567 (2007);


K. S. Virbhadra, and C. R. Keeton, Phys. Rev. D {\bf 77}, 124014
(2008);

K. Hioki, and U. Miyamoto, Phys. Rev. D {\bf 78}, 044007,  (2008);

C. Bambi, and K. Freese, Phys. Rev. D {\bf 79}, 043002 (2009);


V. Perlick, arxiv.org/1010.3416v1[gr-qc];

V. I. Denisov and V. A. Sokolov, J.  Experim. and Theor. Phys., {\bf
113}, 926 (2011);

E. F. Eiroa and C. M. Sendra, Class. Quantum Grav. {\bf 28}, 085008
(2011);

F. Tamburini and B. Thid\'e, arxiv.org/1109.0140v1[gr-qc] ;

L. Amarilla, and E. F. Eiroa, Phys. Rev. D {\bf 85},   064019
(2012);

Z. Stuchlik and M. Kolo\v{s}, J. Cosm. Astropart. Phys. {\bf 10} 008
(2012);

S. Hod, Phys. Rev. D {\bf 87},  024037 (2013).

 E. F. Eiroa and C. M.
Sendra, Phys. Rev. D {\bf 86}, 083009 (2012);

L. Amarilla and E. F. Eiroa, Phys. Rev. D {\bf 87}, 044057 (2013);

C. Ding, C. Liu, Y. Xiao, L. Jiang, R.-G. Cai, Phys. Rev. D {\bf
88}, 104007 (2013);

E.F. Eiroa, C.M. Sendra, Phys. Rev. D {\bf 88}, 103007 (2013);

F. Atamurotov, A. Abdujabbarov, B.  Ahmedov,  Phys. Rev. D {\bf 88},
064004 (2013);

S.-W. Wei and Y.-X. Liu,  J. Cosm. Astropart. Phys. {\bf 11}, 063
(2013);

Z. Stuchl\'ik, A. Kotrlova, and G. T\"or\"ok, Astron. Astrophys.
{\bf 552}, A10 (2013);

J. P. De Andrea,  K. M. Alexander, arxiv.org/1402.5630v2[gr-qc];

C. L\"ammerzahl, J. M\"uller, Gen. Rel. Grav. {\bf 46}, 1 (2014);

 A. Grenzebach, V. Perlick, and
C. L\"ammerzahl, arxiv.org/1403.5234v1[gr-qc].



}



\bibitem[\protect\citeauthoryear{Horowitz \& Myers}{1998}]{Horowitz_98}
G. T. Horowitz, R.C. Myers,  Phys. Rev. D  {\bf 59}, 026005 (1998);

E. Witten, Adv. Theor. Math.  Phys.  {\bf 2}, 6 (1998);

V. Balasubramanian, P. Kraus, Commun. Math. Phys. {\bf 208}, 413
(1999);

 O. Aharony, S. S. Gubser, J.
Maldacena, H. Ooguri, Y. Oz, Phys. Rep. {\bf 323} 183 (2000);

J. M. Cline, J. Descheneau, M. Giovannini and J. Vinet, J. High
Energy Phys. {\bf 06}, 048 (2003).



\bibitem[\protect\citeauthoryear{Susskind \& Lindesay}{2005}]{Susskind_05}
L. Susskind, J. Lindesay, An Introduction to  Black Holes,
Information and the String Theory Revolution: The Holographic
Universe, World Scientific, 2005.


\bibitem[\protect\citeauthoryear{Carter}{1973}]{Carter_73}
B. Carter, in {\it Black Holes, Les Astres Occlus}, ed. by C. de
Witt, B.S. de Witt,
 p.~57 (Gordon and Breach Science Publishers, 1973).

\bibitem[\protect\citeauthoryear{Carter}{1968}]{Carter_68}
B. Carter, Phys. Rev.  {\bf 174}, 1559 (1968).

\bibitem[\protect\citeauthoryear{Stuchl\'ik}{1983}]{Stuchlik_83}
Z. Stuchlik, Bull. Astron. Inst. Czechoslov.  {\bf 34}, 129 (1983).

\bibitem[\protect\citeauthoryear{Misner, Thorne and Wheeler}{1973}]{MTW}
C. Misner, K. Thorne and J.A. Wheeler, {\it Gravitation},  (W.H.
Freeman and Company, San Francesco,  1973).


\bibitem[\protect\citeauthoryear{Wald}{1984}]{Wald_84}
R. M. Wald,    {\it General Relativity}, (The Chicago University
Press, 1984).






\bibitem[\protect\citeauthoryear{Zakharov}{1986}]{Zakharov_86}
A.F. Zakharov,   Sov. Phys. JETP {\bf 64}, 1 (1986).




\bibitem[\protect\citeauthoryear{Darwin}{1958}]{Darwin_58}
C. G. Darwin,   Proc. Roy. Soc. (London)  A {\bf 249}, 180 (1958);

C. G. Darwin,   Proc. Roy. Soc. (London)  A {\bf 263}, 39 (1961).

\bibitem[\protect\citeauthoryear{Kostrikin}{1982}]{Kostrikin_82}
A.I. Kostrikin, {\it Introduction to Algebra}, (Springer, 1982).

\bibitem[\protect\citeauthoryear{Zakharov}{1988}]{Zakharov_88}
A.F. Zakharov,   Sov. Astron. {\bf 32}, 456 (1988).

\bibitem[\protect\citeauthoryear{Zakharov}{1991}]{Zakharov_91}
A.F. Zakharov,   Sov. Astron. {\bf 35}, 147 (1991).

\bibitem[\protect\citeauthoryear{Zakharov}{1994}]{Zakharov_94b}
A.F. Zakharov,    Class. Quant. Grav. {\bf 11}, 1027 (1994).


\bibitem[\protect\citeauthoryear{Lightman et al.}{1975}]{Lightman_75}
A. P. Lightman, W.H. Press, R.H. Price, S.A. Teukolsky,  {\it
Problem Book in
 Relativity and Gravitation},  (Princeton University Press, Princeton, New Jersey, 1975).








\bibitem[\protect\citeauthoryear{Reid et al.}{2014}]{Reid_14}
{M. J. Reid, K. M. Menten, A. Brunthaler et al.,
Astrophys. J. {\bf 783}, 130 (2014).}



\bibitem[\protect\citeauthoryear{Gillessen_09}{2009}]{Gillessen_09}
A. M. Ghez, S. Salim, N. N. Weinberg, J. R. Lu, T. Do, J. K. Dunn,
K. Matthews, M. R. Morris, S. Yelda, E. E. Becklin, T. Kremenek, M.
Milosavljevi\'c, and J. Naiman, Astrophys. J. {\bf 689}, 1044
(2008);

S. Gillessen, F. Eisenhauer, T. K. Fritz, H. Bartko, K. Dodds-Eden,
O. Pfuhl, T. Ott, and R. Genzel, Astrophys. J. {\bf 707}, L114
(2009).

\bibitem[\protect\citeauthoryear{Falcke \& Markoff}{2013}]{Falcke_13}
{H. Falcke, and S.B. Markoff, Class. and Quant. Grav. {\bf 30},
244003 (2013).}


\bibitem[\protect\citeauthoryear{Fish et al.}{2011}]{Fish_11}
V. L. Fish et al., Astrophys. J. Lett. {\bf 727}, L36 (2011).

\bibitem[\protect\citeauthoryear{Doeleman et al.}{2012}]{Doeleman_12}
S. S. Doeleman et al., Science {\bf 338}, 355 (2012).

\bibitem[\protect\citeauthoryear{Blakeslee et al.}{2009}]{Blakeslee_09}
J. Blakeslee et al., Astrophys. J. {\bf 694}, 556 (2009).

\bibitem[\protect\citeauthoryear{Gerbhardt et al.}{2011}]{Gerbhardt_11}
K. Gebhardt et al., Astrophys. J. {\bf 729}, 119 (2011).


\bibitem[\protect\citeauthoryear{Stornaiolo}{2002}]{Stornaiolo_02}
C. Stornaiolo, Gen. Rel. Grav. {\bf 34}, 2089 (2002);

M. Serpico, R. D'Abrusco, G. Longo, C. Stornaiolo, Gen. Rel. Grav.
{\bf 39}, 1551 (2007).

\bibitem[\protect\citeauthoryear{Kagramanova}{2006}]{Kagramanova_06}
V. Kagramanova, J. Kunz, C. L\"ammerzahl, Phys. Lett. B {\bf 634},
465 (2006);

Ph. Jetzer and M. Sereno, Phys. Rev. D {\bf 73}, 044015 (2006);

M. Sereno and Ph. Jetzer, Phys. Rev. D {\bf 73}, 063004 (2006).


\bibitem[\protect\citeauthoryear{Kerr et al.}{2003}]{Kerr_03}
A.W. Kerr, J.C. Hauck, B. Mashhoon, Class. Quantum Grav. {\bf 20},
2727 (2003).

\bibitem[\protect\citeauthoryear{Turyshev et al.}{2009}]{Turyshev_09}
V. T. Toth and S. G. Turyshev, Phys. Rev. D {\bf 79}, 043011 (2009);

 S. G. Turyshev, V. T. Toth, J. Ellis, and C. B. Markwardt,
Phys. Rev. Lett. {\bf 107}, 081103 (2011);

S. G. Turyshev, V. T. Toth, G. Kinsella et al., Phys. Rev. Lett.
{\bf 108}, 241101 (2012).

\end{thebibliography}
\end{document}